\documentclass[aps,prb,preprint,amsmath,amssymb,superscriptaddress, showkeys]{revtex4-2}
\usepackage{hyperref}
\usepackage{longtable}
\usepackage{graphicx}
\usepackage{dcolumn}
\usepackage{bm}
\usepackage{color}
\usepackage{soul} 
\usepackage{amsmath}
\usepackage{nicefrac}

\graphicspath{{./Figures/}}

\newcommand{\NLCWO}{NaLaCoWO$_{6}$}  

\begin{document}

\title{Hybrid improper dipolar density wave in \NLCWO}

\author{Andrea Griesi}\email[corresponding author:]{andrea.griesi@iit.it}
\affiliation{Department of Chemistry, Life Sciences and Environmental Sustainability, University of Parma, Parco Area delle Scienze 17/A, 43124 Parma, Italy}
\affiliation{Istituto Italiano di Tecnologia, Center for Materials Interfaces, Electron Crystallography, Viale Rinaldo Piaggio 34, Pontedera, Italy.}

\author{Enrico Mugnaioli}
\affiliation{Istituto Italiano di Tecnologia, Center for Materials Interfaces, Electron Crystallography, Viale Rinaldo Piaggio 34, Pontedera, Italy.}

\author{Arianna E. Lanza}
\affiliation{Istituto Italiano di Tecnologia, Center for Materials Interfaces, Electron Crystallography, Viale Rinaldo Piaggio 34, Pontedera, Italy.}

\author{Valentina Vit}
\affiliation{Department of Chemistry, Life Sciences and Environmental Sustainability, University of Parma, Parco Area delle Scienze 17/A, 43124 Parma, Italy}

\author{Lara Righi}
\affiliation{Department of Chemistry, Life Sciences and Environmental Sustainability, University of Parma, Parco Area delle Scienze 17/A, 43124 Parma, Italy}

\author{Mauro Gemmi}
\affiliation{Istituto Italiano di Tecnologia, Center for Materials Interfaces, Electron Crystallography, Viale Rinaldo Piaggio 34, Pontedera, Italy.}

\author{Fabio Orlandi}\email[corresponding author:]{fabio.orlandi@stfc.ac.uk}
\affiliation{ISIS Facility, Rutherford Appleton Laboratory - STFC, OX11 0QX, Chilton, Didcot United Kingdom}

\date{\today}

\begin{abstract}
Hybrid Improper Ferroelectricity (HIF) allows the generation of an electrical polarization in the AA'BB'O$_6$ double perovskite materials thanks to the combination of two non-polar octahedral distortions. Nevertheless, for selected combination of the A/A' cations a non-polar incommensurate phase is observed with average symmetry $C2/m$. Thanks to a detailed crystallographic description of the incommensurate phase, based on electron, neutron and x-ray diffraction data, we show that the incommensurate modulation is related to an abrupt change of the out-of-phase tilting along the a- and c-axis whereas the tilting along the b-axis remain constant across the structure. By using group theory and symmetry analysis we show that we observe an incommensurate analog of  HIF which induces a hybrid improper dipolar density wave in \NLCWO. The dipolar ordering is due also in this case to a trilinear invariant involving the commensurate and incommensurate octahedra tilting's.
\end{abstract}

\keywords{}

\maketitle

\section{Introduction}

Concomitant magnetic and electric dipolar ordering represents an intriguing playground for exotic functional properties ruled by the mutual interaction of periodically oriented dipoles.\cite{Eerenstein2006} Such a rare combination of ferroic orders offers the promising opportunity to exploit the magnetoelectric effect and it is commonly obtained by inducing electric polarizability in magnetic materials. Perovskites are one of the most exploited class of materials to obtain such states and the traditional strategies to ensure the breaking of inversion symmetry rely on the incorporation of either \textit{d$^{0}$} metals in the perovskite B-site, involving second-order Jahn-Teller distortions,\cite{Hill2000} or of lone-pair active ions like Bi$^{3+}$ or Pb$^{2+}$ in the structure A-site\cite{Orlandi2014,Catalan2009,Spaldin2019} or by using complex magnetic ordering as for example cycloidal or spiral structures.\cite{Cheong2007}

Recently, the developing of unconventional routes to obtain ferroelectric phases has triggered a considerable interest in the research community.\cite{,Zhao2020,Khalyavin2020,Stroppa2013,Zhu2021,Zhu2020,Yoshida2018,Fukushima2011,Zuo2017,Benedek2011,Orlandi2021,Li2017} These new routes involve the combination of two or more non-polar lattice distortions, like for example octahedral tilting and/or cation/anion orderings, to generate an electrical polarization as a secondary order parameter. The advantage of these strategies derives from the fact that the required non-polar distortions are more common and they are not mutually exclusive with the presence of magnetic cations.\cite{Hill2000} The main disadvantage is related to the magnitude of the polarization that, being a secondary order parameter, is usually small with values of few $\mu C /cm^{-2}$ \cite{Liu2019,Liu2015,Khalyavin2020,Orlandi2016}. Moreover, the switching mechanism might involve a quite complex energy landscape since, in order to switch the polarization, only one of the two (or more) antipolar distortions needs to change sign leaving the others invariant. \cite{Benedek2011,Huang2016,Li2020,Liu2019} Nevertheless, these new mechanisms to achieve polar states have led to new exciting dipolar ordered phases such as the ferrochiral ordering in Ba(TiO)Cu$_4$(PO$_4$)$_4$\cite{Hayashida2021} and the incommensurate helical dipole ordering in BiCu$_{0.1}$Mn$_{6.9}$O$_{12}$,\cite{Khalyavin2020} which open up the possibilities of complex non-collinear textures like polar skyrmions recently observed in (PbTiO$_3$)$_n$/(SrTiO$_3$)$_n$ superlattices.\cite{Das2019}

\begin{figure}
\includegraphics[width=0.45\textwidth]{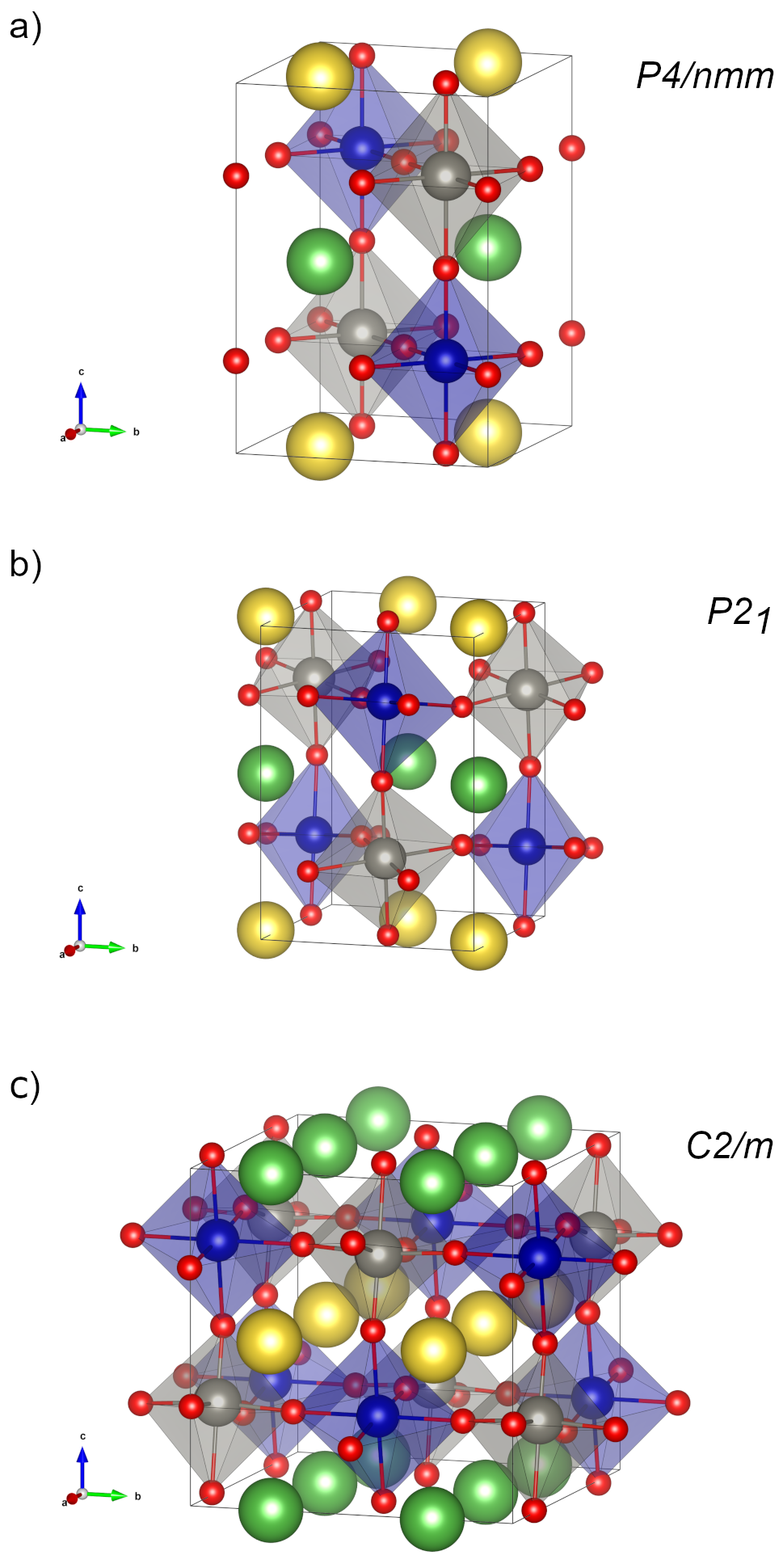}%
\caption{\label{1} \textbf{General structure of double ordered perovskites.} a) parent $P4/nmm$ structure, b) polar $P2_1$ structure with $a^-a^-c^+$ octahedral tilting and c) $C2/m$ structure with the $a^0b^-c^0$ tilting scheme. The A/A' cations are shown as yellow and green spheres, whereas the B/B' cation as gray and blue sphere and the oxygen anion as red spheres.}
\end{figure}

“Hybrid Improper Ferroelectricity” (HIF) is one of these new routes to induce a spontaneous ferroelectric polarization.\cite{Young2015,Stroppa2013,Zhu2020,Yoshida2018,Fukushima2011,Benedek2011,Zuo2017} HIF in perovskite-based materials arises from the coupling of two non-polar modes, related to the tilt of the octahedral coordination units, which induce a polar mode as secondary order parameter. The coupling in the system free energy is a trilinear invariant between the order parameters describing the latter distortions. Several examples of HIF were encountered in layered perovskite compounds\cite{Zhu2021,Zhu2020,Yoshida2018,PerezMato2004,Sun2016} as well as in double perovskites (with general formula AA'BB'O$_6$) with ordered distribution of heterovalent transition metals in the A site.\cite{Fukushima2011,Zuo2017,Shankar2020,Shankar2021} Such systems, whose parent structure is depicted in Figure \ref{1}a, have been experimentally obtained with a large variety of A/A' and B/B' combinations.\cite{Lopez1994,Zuo2017,Shankar2020,Shankar2021,Garcia2012,Zuo2019,Arillo1997,De2018,Borchani2017} The B and B' cations are often W$^{6+}$ (or another second-order Jahn-Teller active metal) combined with \textit{3d} transition metals like Mn, Fe, Co and Ni.\cite{Lopez1994,Zuo2017,Shankar2020,Shankar2021,Garcia2012,Zuo2019,Arillo1997,De2018} These systems display magnetic ordering at low temperature, usually below 15 K, due to the weak B-O-W-O-B super-superexchange interactions.\cite{Shankar2020,Shankar2021,De2018}

In particular, some compounds in the NaA'BWO$_6$ family are characterized by the polar monoclinic $P2_1$ symmetry with the $a^-a^-c^+$ tilting scheme, which allows HIF.\cite{Fukushima2011,Shankar2020,Shankar2021,De2018} Computational works and symmetry considerations indicate that cationic ordering in A and/or B sites is one of the most efficient chemical tool to enable a stable ferroelectric polarization via the HIF mechanism.\cite{Stroppa2013,Benedek2011,Young2015,Shankar2020} Young \textit{et al.}\cite{Young2015} suggested that the key factor defining the electric polarizability in AA'MnWO$_6$ is the large ionic radii difference between A and A', confirming that the electric polarization is associated with HIF and it is promoted by the large octahedral tilting influenced by the smaller size of the A' cations.

On the contrary, other AA'BWO$_6$ compounds crystallize in the centrosymmetric $C2/m$ space group (Figure \ref{1}c), exhibiting the tilting scheme $a^0b^-c^0$. Compositions based on the A/A' couple Na/La feature the C-centered crystal structure at ambient conditions and transform into the $P2_1$ acentric phase at lower temperature with a first order transition characterized by a very large thermal hysteresis.\cite{Zuo2019,Zuo2017} Electron diffraction (ED) and High Resolution Electron Microscopy (HREM) investigations revealed that the monoclinic C-centered crystal structure in systems like NaLaMgWO$_6$,\cite{Garcia2008,King2009} NaLaCoWO$_6$,\cite{Zuo2019} KLaMnWO$_6$\cite{Garcia2011} and NaCeMnWO$_6$\cite{Garcia2012} manifests incommensurate structural modulations. HREM measurements revealed an unusual nano-checkerboard contrast, corresponding to incommensurately modulated domains.\cite{Garcia2011,Garcia2011} The incommensurate structure has been related to a compositional modulation related to an inhomogeneous A/A' distribution and the complex contrast in HREM has been interpreted as originated by the regular distribution of perovskite regions showing a different content of Na and La.\cite{Garcia2011} Other studies, instead, associate the incommensurate modulation to displacive topological distortions of the octahedral tilting and periodic twinning.\cite{Garcia2012} The origin of the complex structural modulation in the ordered AA'BWO$_6$ perovskite is controversial and the crystallographic description of the incommensurate modulated structure is still to be fully described. 

For these reasons, we synthesized the NaLaCoWO$_6$ double perovskite and we investigated its incommensurate modulated structure by 3D electron diffraction (3D ED),\cite{Gemmi2019,Lanza2019,Steciuk2019,Steciuk2020} neutron and x-ray powder diffraction. 3D ED has the ability to work on single nano-crystalline domains, allowing the determination of the modulation vector $\stackrel{\rightarrow}{q}$ and of the modulated structure by applying the superspace formalism.\cite{vanSmaalen2004,Lanza2019,Steciuk2019,Steciuk2020} The final structure refinement has been performed with the combination of neutron and x-ray diffraction data, allowing an accurate description of the modulated structural distortions and their couplings. Our findings and detailed symmetry analysis suggest an incommensurate equivalent of HIF, which leads to the emergence of a hybrid improper dipolar density wave.

\section{Methods}

The synthesis of \NLCWO{} was performed by solid state reaction mixing Na$_2$CO$_3$, La$_2$O$_3$ and CoWO$_4$, in stoichiometric amounts. The CoWO$_4$ precursor was prepared using stoichiometric quantities of CoO and WO$_3$ and heat treated at 1173 K for 8 hours. The reagent oxides were mixed and heat treated at 1198 K in air for 6 hours with a heating ramp of 4.5 K/min. Afterward, the sample was grinded and pressed to form a pellet and sintered at a temperature of 1223 K with a ramp of 2.5 K/min for 8 hours. The sample purity was checked by x-ray diffraction and small amounts of LaNaW$_2$O$_8$ (3.6(1) w\%), CoO (1.04(3) w\%) and Co$_3$O$_4$ (4.7(1) w\%) were observed.

Precession-assisted 3D ED and energy-dispersive x-ray (EDX) spectroscopy were performed with a Zeiss Libra 120 transmission electron microscope working at 120 kV and equipped with a LaB$_6$ thermionic source and a Bruker EDX XFlash6T-60 detector for EDX analysis. The samples for electron diffraction were ground in an agate mortar and the resulting powder was suspended and sonicated in ethanol and drop-deposited on a carbon-coated copper grid. Sequential electron diffraction patterns were collected in nanodiffraction mode with a parallel beam of 150 nm, obtained by selecting a condenser aperture of 5 $\mu$m. The diffracted beams were energy-filtered on the zero-loss peak with an in-column $\omega$-filter and a slit width of about 20 eV.\cite{Gemmi2013} Patterns were collected in stepwise mode with a protocol similar to the one proposed by Mugnaioli et al.\cite{Mugnaioli2009} The precession motion of the electron beam around the optical axis was obtained by a Nanomegas Digistar P1000 device.\cite{Vincent1994} The semi-aperture of the precession cone was set to 1$^{\circ}$. 3D ED data was collected over a tilt range of up to 120$^{\circ}$, in steps of 1$^{\circ}$. After each tilt step, a STEM image was taken to check the crystal movement and the electron beam was positioned back in the same point of the crystal. Diffraction patterns were recorded by an ASI Timepix single-electron detector.\cite{Nederlof2013} The program PETS 2.0 was used for the determination of the unit cell and modulation vector and for the extraction of the electron diffraction intensities on data from thirty different nanocrystals.\cite{Palatinus2019} 

The x-ray diffraction (XRD) measurements were performed with a STOE Stadi P diffractometer equipped with Cu--K$\alpha_1$ radiation, a STOE\&Cie Johansson monochromator and a Dectris MYTHEN2 1K detector. The samples were ground and prepared as thin disks sandwiched by acetate foils to perform  measurements in transmission mode. 

Time-of-flight neutron powder diffraction data (TOF-NPD) were collected on the cold neutron diffractometer WISH at the ISIS second target station (UK).\cite{Chapon2011} The data were focused on 4 detector banks with average $2\theta$ of 154, 121, 90 and 57 $^{\circ}$, each covering 32 $^{\circ}$ of the scattering plane.

All structure refinements were performed with the Jana2006 software.\cite{Petricek2014} The 3D ED data has been treated within the kinematical approximation. Group theory and symmetry analysis calculations have been performed with the help of the ISOTROPY and ISODISTORT software.\cite{Campbell2006,Hatch2003} The free energy invariants have been calculated using the ISOTROPY and INVARIANT software.\cite{Campbell2006,Hatch2003} The crystallographic structures are drawn with the help of the VESTA software.\cite{Momma2011}

\section{Result and Discussion}

\subsection{Structure solution and refinement}

\begin{figure*}
\includegraphics[width=0.95\textwidth]{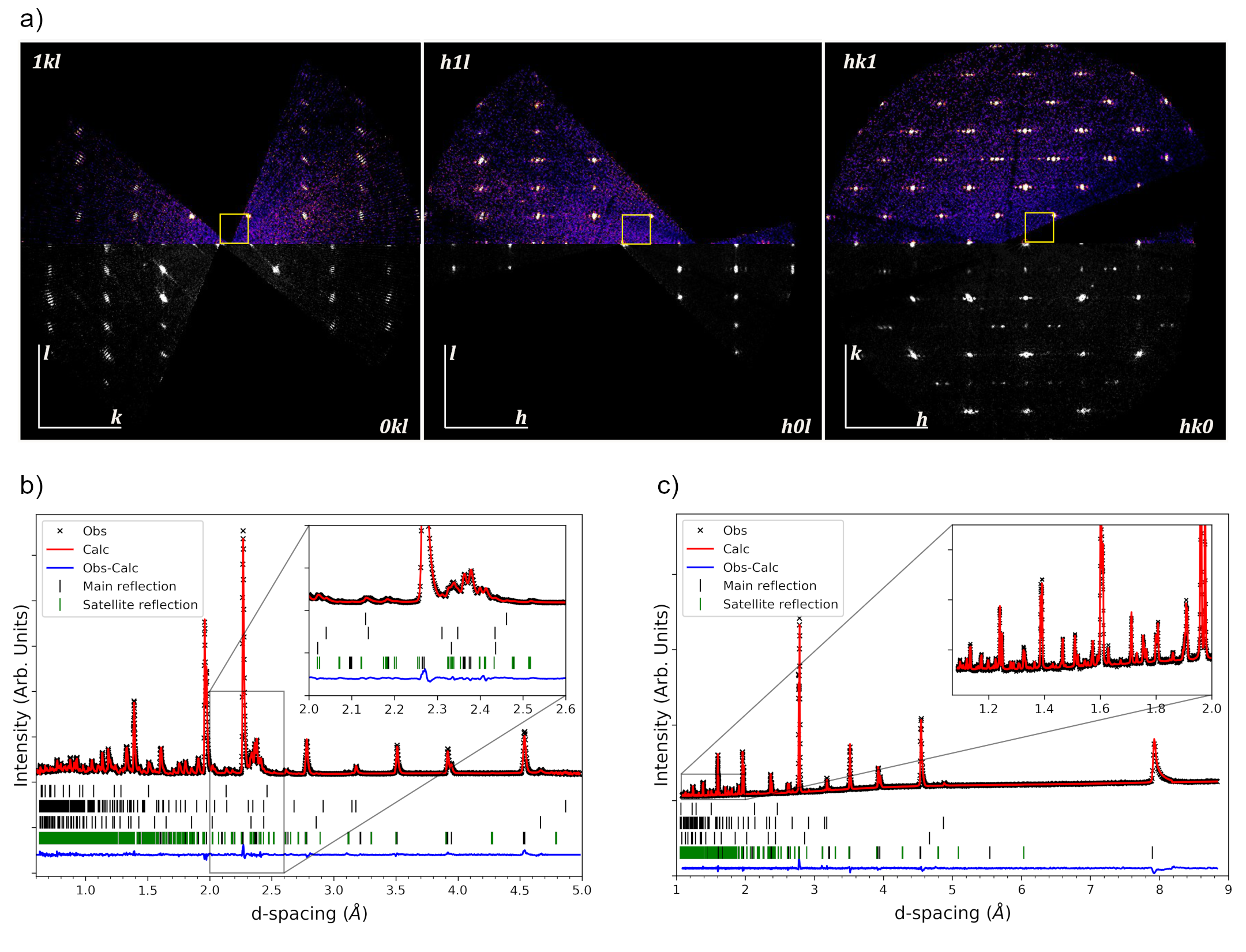}%
\caption{\label{2} \textbf{Diffraction data collected on \NLCWO.} a) Sections of the \NLCWO{} reciprocal space obtained from the 3D ED data showing the modulation along the $a^*$ direction. The black areas, devoid of reflections, are areas of the reciprocal space that are not measured due to the geometrical constrain of the experiment. b) Rietveld plot of the TOF-NPD data collected on WISH at average $2\theta$ of 154$^{\circ}$ and c) Rietveld plot of the XRD data converted in d-spacing for an easier comparison with the neutron data. Observed data (black crosses), calculated pattern (red line) and difference (blue line) are reported. The bottom black tick marks indicate the Bragg position of the main phase and the green tick marks indicate the satellite reflections. The upper three rows of tick marks indicate the Bragg position of CoO, LaNaW$_2$O$_8$ and Co$_3$O$_4$ from top to bottom.}
\end{figure*}

3D ED data collected on \NLCWO{} nanocrystals (see Figure S1)\cite{SUP} clearly show that the main reflections are accompanied by first order satellites (Figure \ref{2}a). The main reflections can be indexed with a pseudo-cubic cell with: a=7.83(3) \AA, b=7.84(2) \AA, c=7.89(8) \AA{} and $\beta$ close to 90$^{\circ}$ (the actual value refined against XRD data is 90.160(10)$^{\circ}$). From the reconstruction of the reciprocal space, the extinction rule h+k=2n for the main reflections can be easily determined, indicating a C-centered lattice with no other extinctions observed. A reasonable model, with $C2/m$ symmetry, was found \textit{ab-initio} using the charge flipping algorithm\cite{Palatinus2013} on the 3D ED data. The average structure of the double ordered perovskite is in agreement with the one reported in literature,\cite{Zuo2019} with rock salt ordering of the CoO$_6$ and WO$_6$ octahedra, layer ordering of the Na and La cations and an average tilting pattern $a^0b^-c^0$. 

Satellite reflections are observed in the 3D ED data along the $a^*$ direction, with a modulation vector q=0.172(2)$a^*$ . Some crystals show a strong twinning, which exchanges the $a^*$ and $b^*$ directions (see figure S2\cite{SUP}). The collection of 3D ED data on a single crystalline domain allowed to determine the reflection conditions in four index notations as hklm: h+k=2n and h0lm: m=2n, which are consistent only with the superspace group $C2/m(\alpha0\gamma)0s$. A first tentative refinement of the modulated structure against 3D ED data alone returned a structure with large displacive modulation of the La and Na atoms, clearly in contrast with the x-ray diffraction data where the satellite reflections are almost absent (Figure \ref{2}c). In earlier structural studies, aimed to disclose the nature of the incommensurate modulation in \NLCWO, it was suggested that the incommensurate distortions involve mainly the oxygen positions.\cite{Zuo2019} The authors indicated the presence of monoclinic twinned domains characterized by the inversion of the octahedral tilting, without providing a crystallographic description of such twinning.

\begin{figure*}
\includegraphics[width=0.92\textwidth]{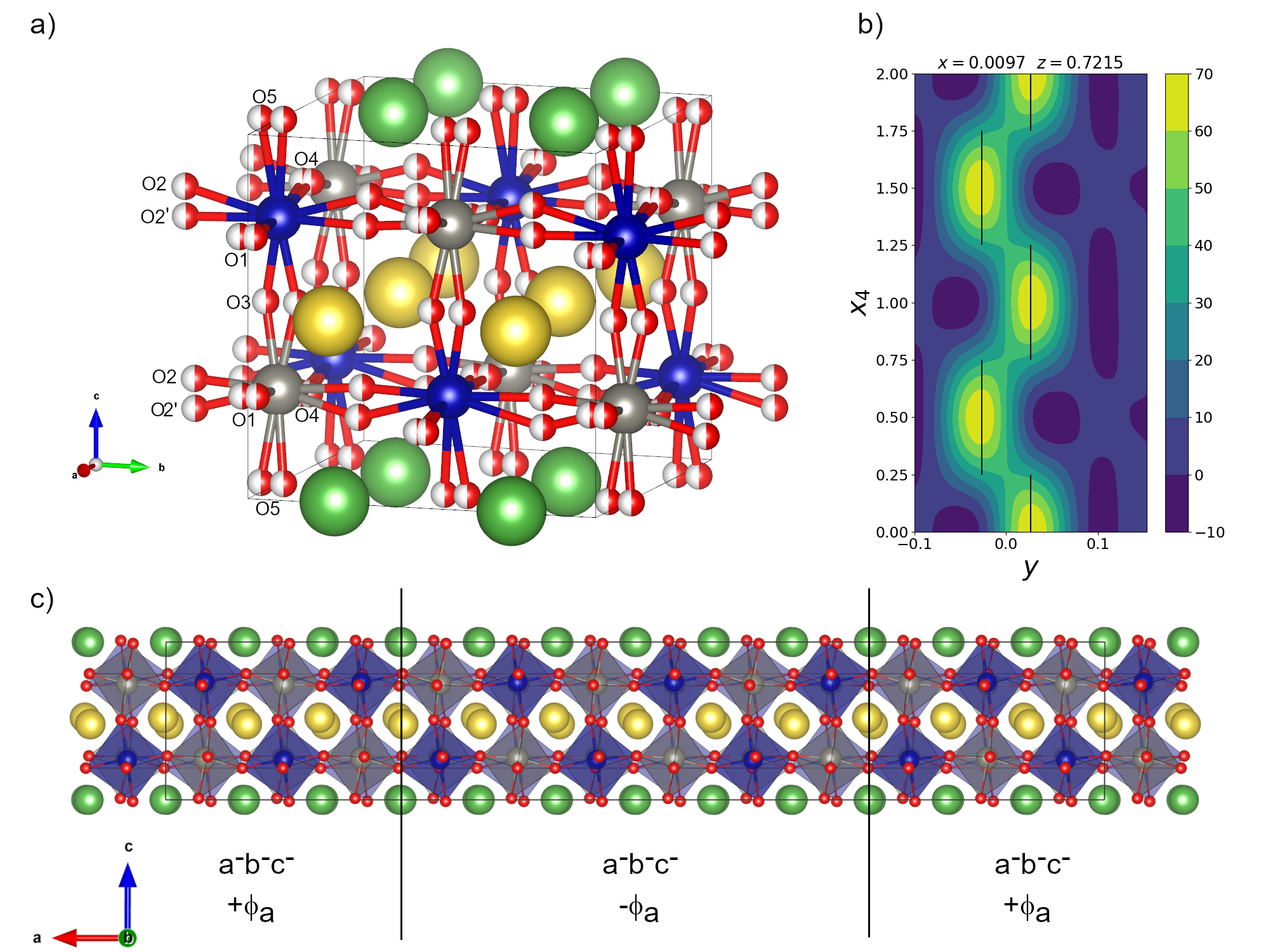}%
\caption{\label{3} \textbf{Modulated structure of \NLCWO.} a) Average structure showing the disorder of the BO$_6$ polyhedra. Only the disorder of the oxygen positions is displayed for clarity. b) Fourier map (in Arb. Units) calculated around the O1 position. The black line in the map corresponds in 3D space to O1 in position (x, y, z) and in the symmetry-generated position (x, --y, z). The map is showing the discontinuous occupancy of the split positions along the $x_4$ direction. $x_4$ is the internal coordinate defined as $t+\stackrel{\rightarrow}{q} \cdot \stackrel{\rightarrow}{r}$ where t is the initial phase of the modulation function, $\stackrel{\rightarrow}{q}$ is the modulation vector and $\stackrel{\rightarrow}{r}$ is a position vector (x,y,z)\cite{vanSmaalen2004,Petricek2014}. c) Evolution of the incommensurate structure across 8 unit cell along the a direction. The vertical lines indicate the region where the tilting angle along the a and c direction changes sign. In all panels the Na and La cations are shown as yellow and green spheres respectively, whereas the W and Co cations as gray and blue sphere and the oxygen anion as red spheres}
\end{figure*}

To disclose the nature of the incommensurate distortions, high-resolution powder neutron diffraction, which is more sensitive to light ions, was carried out at room temperature. The simultaneous refinement of the average structure against the TOF-NPD data and the powder x-ray data results in a good fit. The average structural model is featured by elongated anisotropic ADPs for the oxygen positions suggesting a possible disorder or splitting of these sites. This possibility has been tested against the data with a clear improvement of the refinement and a normalization of the thermal parameters. The splitting of the equatorial oxygen (O1, O3, O4 and O5 see figure \ref{3}a) does not require additional atomic positions but can be obtained by shifting the ion from the 4i Wyckoff position sitting on the mirror plane. On the contrary, the splitting of the apical O2 site has been model with two symmetry-independent positions (O2 and O2' see table S1\cite{SUP}). The average splitting distance is $\approx$ 0.5 \AA. A similar splitting, even if considerably smaller, has been observed for the cations in the structure. The disordered anion positions result in two possible octahedral orientations surrounding the Co$^{2+}$ and W$^{6+}$ cations. This suggests that two different tilting arrangements of the octahedral frameworks coexist in the crystal structure of \NLCWO{}, likely determining the incommensurate modulation. 

To confirm this hypothesis, an incommensurate model was conceived starting from the unit cell, the modulation vector and the superspace group symmetry ($C2/m(\alpha0\gamma)0s$) determined by 3D ED. In the structural refinement, based on TOF-NPD and XRD data, the modulation of the split oxygen positions has been described with the use of discontinuous crenel functions,\cite{EVAIN2004} avoiding that in any point of the crystal both positions are simultaneously occupied. For the equatorial oxygen position (O1, O3, O4 and O5), to fulfill the latter requirement is sufficient to define a single crenel function. Indeed, the presence in the superspace group of the symmetry element $\{m_y|000s\}$ assures that only one position above or below the mirror plane is occupied throughout the crystal (see Figure \ref{3}b). On the contrary, for the two apical O2 and O2', the absence of this symmetry element obliges to constrain the origin of the crenel function, x$_{4,0}$ ($x_4$ is the internal coordinate defined as $t+\stackrel{\rightarrow}{q} \cdot \stackrel{\rightarrow}{r}$ where t is the initial phase of the modulation function, $\stackrel{\rightarrow}{q}$ is the modulation vector and $\stackrel{\rightarrow}{r}$ is a position vector (x,y,z))\cite{vanSmaalen2004,Petricek2014}, for the two apical O2 and O2' positions to have a 0.5 difference to fulfill the same physical requirement. A similar modulation function has been applied to the cation positions.

This model returns optimal agreement factors (see table S1)\cite{SUP} for the simultaneous fitting of the powder data (TOF-NPD and XRD) as shown in Figure \ref{2}b-c. It is worth to underline that the superspace group allows the crenel origin parameter (x$_{4,0}$) for all positions to possess any value in the 0-0.5 range, nevertheless the refinement strongly suggests that the x$_{4,0}$ value for all atoms (except for the O2' which is constrained to be the O2 value +0.5) to be zero within the uncertainty of the refinement. This is further confirmed by the absences of the 1111 and 110$\overline{1}$ reflections from the neutron pattern (indeed simulations with x$_{4,0}$ values different from 0 returns strong peaks at these positions). For this reason, the x$_{4,0}$ parameters were fixed to zero in the final refinement. It is worth to underline that the incommensurate model here proposed is very simple and the only refinable parameters, which influence the satellite intensities, are the positions of the split atoms in the average structure. To check if the refined model is compatible with 3D ED data, the incommensurate model obtained from the powder diffraction data has been refined against 3D ED data within the kinematical approximation. The refinement is stable and converges with agreement factor of wR=18.49\%, R=17.59\% and GOF=11.07. The final results of both refinement (3D ED data and powder neutron/x-ray data) are reported in table S1\cite{SUP} showing good agreement. 

The refined modulated structure is shown in Figure \ref{3}c. The incommensurate modulation can be described by an abrupt change of the tilting angle $\Phi$ along the a and c directions which resembles a twinning boundary. The tilting pattern in the structure remain constant as $a^-b^-c^-$, with the tilting angle $\Phi$ along the a and c direction abruptly changing its value but maintaining its antiphase character. Only at the boundary between the regions with $+\Phi$ and $-\Phi$ the tilting pattern is $a^+b^-c^+$. On the contrary, the value of the tilting angle along the b direction remain constant across the structure. The antiphase tilting along the c direction is a direct consequence of the x$_{4,0}$ parameter being 0 and we will see in section III-B its importance in the emergence of the dipolar density wave. The anion distortions are responsible for the modulation, which explains the weakness of satellite peaks in the XRD compared with NPD, and the smaller cations distortion follows the changes in octahedral tilting.

The crystallographic description of the modulated structure enables revisiting previous interpretation of this complex distortion. As remarked before, the structural modulation observed in systems like NaLaBB'O$_6$ has been interpreted in terms of irregular distribution of the A and A' ion to form alternated off-stoichiometric domains composing the characteristic checkerboard contrast.\cite{Garcia2008,Licurse2013} In the combined XRD/TOF-NPD structural refinements, the La and Na sites resulted fully occupied and neither mixing of the two heterovalent ions nor randomly distributed vacancies are observed. Furthermore, the identified impurities are not in contrast with the occurrence of the 1:1 ratio of the La and Na in the doubly ordered perovskite. These assessments are, in general, supported by EDX analysis (see figure S3)\cite{SUP}, even if this measurement would suggests a small amount of Na vacancies in selected regions of the doubly ordered perovskite. This can be ascribed to Na evaporation induced by the electron irradiation documented in literature by the observation of the segregation of metallic Na and Na$_2$O.\cite{Egerton2019,Jiang2008} 

\begin{figure*}
\includegraphics[width=0.95\textwidth]{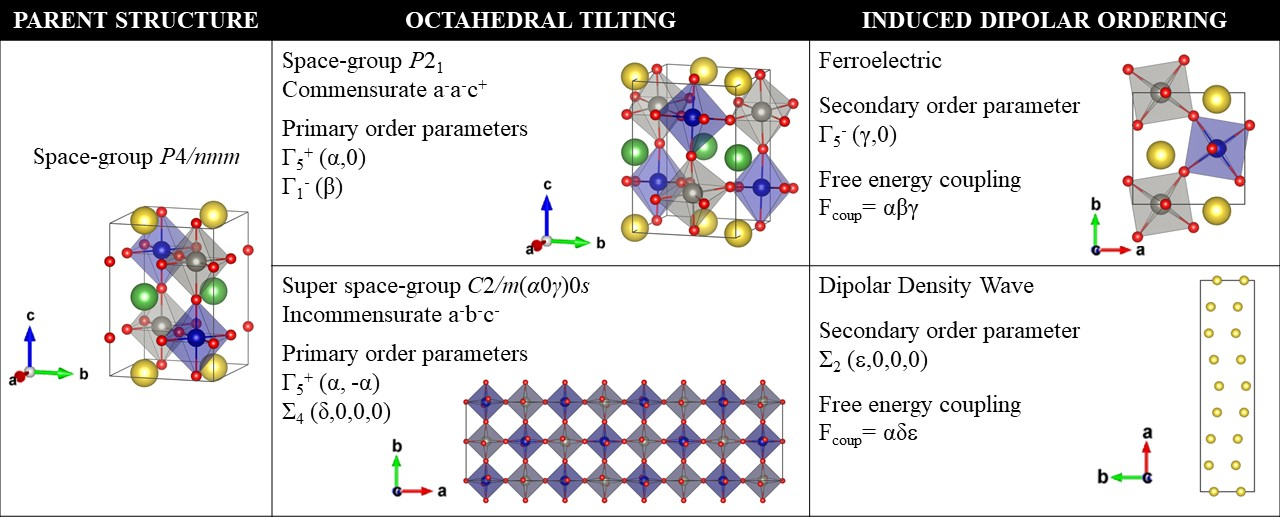}%
\caption{\label{4} \textbf{Symmetry analysis of the $P2_1$ and $C2/m(\alpha0\gamma)0s$ structures.} Schematic display of the primary octahedral tilting distortions and of the corresponding induced dipolar ordering for the phases discussed in the main text.}
\end{figure*}

Interestingly, the characteristic checkerboard contrast in the HREM images, which is usually interpreted as a cation ordering effect, has been explained as the occurrence of incommensurate oxygen displacement distortions in in Li$_{1-x}$Nd$_{2/3-x}$TiO$_3$.\cite{Abakumov2013} These displacements create region with different TiO$_6$ tilting patterns, which are at the basis of the observed bright and dark contrast in the HREM images.\cite{Abakumov2013} In this regards, the crenel modulations functions used to describe the incommensurate tilting patterns in \NLCWO{} can well describe the alternating region observed in the HREM images in different NaLaBB'O$_6$ systems.\cite{Garcia2008,Garcia2011,Licurse2013}

\subsection{Symmetry analysis and the hybrid improper dipolar density wave}

Before discussing the details of the coupling and of the dipolar ordering observed in the incommensurate phase of \NLCWO, it is worth to refresh the HIF mechanism observed in the commensurate $P2_1$ phase. In the remaining of the paper we will describe the distortion of all phases with respect to the parent $P4/nmm$ structure shown in Figure \ref{1}a. This will simplify the coupling terms and the group theory calculations, nevertheless it is still worth to underline some important points regarding the cation ordering and how this allows HIF in these system. The $P4/nmm$ phase is obtained from the parent simple perovskite structure (s.g. $Pm\overline{3}m$) by the action of the $R_{1}^{+}$ irreducible representation (irrep), which correspond to the chessboard ordering of the B site cations, and by the action of the $X_{3}^{-}$ irrep, which gives rise to the layer ordering on the A-site. The latter is important for HIF, as it removes the inversion symmetry from the B site point group allowing the octahedral tilting's to break the macroscopic spatial inversion symmetry\cite{Fukushima2011}

HIF in the $P2_1$ commensurate phases of the AA'BB'O$_6$ perovskites is strictly related to the $a^-a^-c^+$ octahedral tilting.\cite{Fukushima2011,Shankar2020,Shankar2021,De2018} In particular the out-of-phase tilting along the a and b directions is related to the $\Gamma_{5}^{+}$ irrep with order parameter $(\alpha,0)$, whereas the in-phase tilting around the c axis is due to the $\Gamma_{1}^{-}$  irrep with order parameter $\beta$. These two distortions acting together on the parent $P4/nmm$ structure will induce a secondary polar distortion, which transform as the $\Gamma_{5}^{-}$ irreps with order parameter $(\gamma,0)$, thanks to the trilinear free energy invariant $\alpha\beta\gamma$ (Figure \ref{4}). The polar $\Gamma_{5}^{-}$  distortion is a uniform opposite shift of the cation/anion in the structure that result in a net dipolar moment along the b-axis of the $P2_1$ structure. This induced polarization is switchable by an external electric field and the switching mechanism requires the change in sign of either the $\alpha$ or $\beta$ order parameters which correspond to a sign change of the $a^-a^-c^0$ or $a^0a^0c^+$ tilting respectively.

The scenario in the incommensurate phase of the \NLCWO{} compound is similar but results in an incommensurate ordering of the induced dipoles. The $C2/m(\alpha0\gamma)0s$ phase refined in this work can be obtained from the parent $P4/nmm$ structure by the action of two distortions related also in this case to the octahedral tilting's. The first is the commensurate out-of-phase tilting along the b direction. This transform as the $\Gamma_{5}^{+}$ irrep as for the $a^-a^-c^0$ tilting in the $P2_1$ phase but with a different order parameter direction $(\alpha,-\alpha)$. The second distortion is the incommensurate out-of-phase tilting along the a and c directions which transform as the $\Sigma_{4}$ irrep with order parameter $(\delta,0,0,0)$ (this is a direct consequence of the $x_{4,0}$ parameter of the crenel function being equal to 0). These two distortions acting on the tetragonal parent structure induce, as secondary order parameter, an incommensurate displacement of the cation along the b direction which transform as the $\Sigma_{2}$ irrep $(\epsilon,0,0,0)$ thanks, also in this case, to a trilinear invariant $\alpha\delta\epsilon$ in the free energy as schematically shown in Figure \ref{4}. The analogies between the commensurate $P2_1$ case and the incommensurate $C2/m(\alpha0\gamma)0s$ one let us to conclude that in the latter case we have observed the development of an incommensurate ordering of the induced dipoles with the same propagation vector as the incommensurate tilting. The induced dipoles are oriented along the b axis, in analogy with the $P2_1$ phase, and their amplitude is modulated with the period of the modulation vector, resembling the electric analog of a spin density wave. For the latter reason we describe the incommensurate dipolar ordering as a hybrid improper dipolar density wave.

\section{Conclusion}

We have shown that the $C2/m$ phase observed in the \NLCWO{} compound is a modulated phase with an incommensurate out-of-phase tilting along the a and c direction. The combined use of electron, neutron and x-ray diffraction data allowed the refinement and the description of the incommensurate phase within the superspace formalism. The peculiar combination of the incommensurate tilting along two directions and of the commensurate one along the b direction induce a dipolar density wave in perfect analogy with HIF observed in the $P2_1$ phase. The observation of a dipolar density wave adds another piece to the analogy between electrical and magnetic dipoles\cite{Khalyavin2020} and opens the possibility to observe new peculiar phases, like for example the “devil's staircase” and electric solitons,\cite{Bak1980,Rossat-Mignod1977,Bak1982} which are usually exclusive to magnetic orderings. 

\begin{acknowledgments}

The authors acknowledge the Science and Technology Facility Council (STFC-UKRI) for the provision of neutron beam-time on the WISH diffractometer. F.O. acknowledge fruitful discussion with Dr Pascal Manuel and Dr Dmitry Khalyavin. A.G., E.M., A.L., M.G. would like to thank Regione Toscana for funding the purchase of the ASI MEDIPIX detector through the FELIX project (POR CREO FERS 2014-2020).

\end{acknowledgments}

\end{document}